\documentclass[a4paper,10pt]{article}
\usepackage{amsmath}
\usepackage{amsfonts}
\usepackage{amssymb}
\usepackage{graphicx}
\begin{document}

\title{  \textsc{Graviton mass  constraint from   CMB}}

\author{N. Malsawmtluangi $ \footnote{e-mail: tei.naulak@uohyd.ac.in}$
        and
        P. K. Suresh $\footnote{e-mail: sureshpk@uohyd.ac.in}$ \\[.2cm]
{\small \it School of Physics, University of Hyderabad. }\\
{\small \it P. O. Central University, Hyderabad 500046. India.}}

\date{\empty}

\maketitle

\begin{abstract}
The  effect of primordial massive gravitational waves on the BB-mode correlation angular power spectrum of CMB is  studied for  several inflation models. The     angular power spectrum  with  the  BICEP2/Keck Array and Planck  joint data  suggests  further constraint on the lower and upper bounds on the mass of primordial gravitons
\end{abstract}

\section{Introduction}

The force of gravity is believed to be mediated by  spin-2 particles called gravitons which are  commonly considered as massless. However, initiating   with the idea of a spin-2 particle with non-zero rest mass, several approaches have been taken to introduce mass to  graviton \cite{FP, vDVZ, zak, BD,  vainshtein, GS}. Endowing graviton with mass leads to extra degrees of freedom which do not decouple as  graviton mass approaches to zero implying   that the general relativistic (GR) case cannot be recovered \cite{vDVZ, zak}. Some of the approaches to massive gravity suffer from pathologies like the presence of ghost mode \cite{BD}, discontinuity when the mass approaches to zero limiting case and so on \cite{vainshtein}, and several theories have been proposed to fix these problems and also attempted to formulate a consistent theory of massive gravity \cite{GS, Rubakov, Dubovsky, dRGT, dRGT2, Sgr}. 
At the same time, there have been several attempts to estimate the mass of graviton from astrophysical sources and  primordial gravitational waves (GWs)  \cite{m1, m2, m3, m4, m5, m6, m7}. It is proposed   that if graviton  mass is comparable to the Hubble parameter, then it can provide  a repulsive effect at cosmological distances and  hence lead to  the late  time cosmic acceleration, thereby  suggesting that the massive gravitons  responsible  for  the current accelerating phase of the universe  instead of dark energy. %There  is an interesting proposal that  massive gravitons would  comprise of cold dark matter as well \cite{cdm}. 

  In this paper, we consider the Lorentz-violating massive gravity theory in which the Lorentz invariance is spontaneously broken by a convenient choice of the vacuum for the Goldstone fields, and the mass parameters are chosen in such a way that the pathologies are absent, and the scalar and vector modes behave exactly like those in the general relativistic case. Hence,  the modification of the gravity  comes only from the tensor modes and the dispersion relation of gravitational waves  acquires an effective mass and is relativistic \cite{Rubakov, Dubovsky}. According to this theory, the bound on the primordial graviton mass comes from the exponential decay in the Yukawa potential, putting the upper bound for the graviton mass to be $\le 10^{-30}$ eV \cite{Dubovsky, cmbmg2}. The lower bound for graviton mass has been proposed to be $ 1.239 \times 10^{-32}$ eV  \cite{mgtensor}. The small mass of graviton is expected to have an effect on the temperature anisotropy and polarization spectra of  the cosmic microwave background (CMB)  \cite{cmbmg2, cmbmg1}. The  imprint of  primordial gravitational waves   on CMB anisotropy  can be observed through the angular power spectrum of CMB in the form of  B-mode polarization \cite{marc, marc2, baskaran, lpg}. The  observations of B-mode polarization on CMB would   not only verify the theory of inflation itself but  also help in constraining  many inflation models  \cite{jmart1, jmart2, jmart3}. The detection of B-mode polarization of CMB or the primordial GW itself  would  provide a clear bound on the mass of primordial graviton. Hence, in this paper, we  study the effect of the primordial massive GWs on the BB mode correlation angular power spectrum  of CMB for various inflation models and  the results are compared with  the  recent BICEP2/Keck Array  and Planck  collaboration data \cite{bkp} and thereby  to  obtain  constraint on the  mass of  primordial gravitons.

\section{Massive gravitational waves}
For massive gravity, the action  can be written in terms  of the Einstein-Hilbert action and the Goldstone action as  \cite{Rubakov, Dubovsky},
\begin{eqnarray}\label{action}
S &=& S_{EH} + S_{G}, \nonumber \\ 
 &=&\int d^4x \sqrt{-g}[-m_{pl}^2R + \Lambda^4F(Z^{ij})],
\end{eqnarray}
where $\Lambda$ characterizes the cutoff energy scale for low energy effective theory. $F$ is a function of the Goldstone field, metric components and  its derivatives. The second term in the  above action leads to violation of  the Lorentz symmetry.   It is assumed that ordinary matter field is minimally coupled to the metric.

The argument $Z^{ij}$ can be obtained with the help of the  following expressions
\begin{eqnarray}
Z^{ij} &=& X^{\gamma}W^{ij}, \nonumber \\
X &=& \Lambda^{-4}g^{\mu \nu} \partial_{\mu}\zeta^0\partial_{\nu}\Phi^0, \nonumber \\
W^{ij} &=& \Lambda^{-4}g^{\mu \nu} \partial_{\mu}\Phi^i\partial_{\nu}\Phi^j - \frac{V^iV^j}{X}, \nonumber \\
V^i &=& \Lambda^{-4}g^{\mu \nu} \partial_{\mu}\Phi^0\partial_{\nu}\Phi^i,
\end{eqnarray}
where  $\Phi^0 (x)$, $\Phi^i (x)$, ($i = 1,2,3$) are the four scalar fields and $\gamma$ is considered as a constant free parameter.

For  Eq.(\ref{action}),   the vacuum solutions corresponding to  the  flat Friedmann-Lemaitre-Robertson-Walker (FLRW) metric can be written as
\begin{eqnarray}\label{ug}
g_{\mu \nu} &=& a^2 \eta_{\mu \nu}, \nonumber \\
\Phi^0 &=& \Lambda^2t, \\
\Phi^i &=& \Lambda^2 x^i. \nonumber
\end{eqnarray}
where $a $ is the scale factor  for the FLRW metric and  $\eta_{\mu \nu}$ is the flat space metric. 

The metric  $g_{\mu \nu}$  with perturbations can be written as
\begin{equation}
g_{\mu \nu} = a^2\eta_{\mu \nu} + \delta g_{\mu \nu},
\end{equation}
where  the metric perturbations $\delta g_{\mu \nu}$ are taken   after the spontaneous Lorentz symmetry breaking.

The components of the metric perturbation  are given by
\begin{eqnarray}
\delta g_{00} &=& 2a^2 \varphi, \nonumber \\
\delta g_{0i} &=& a^2 (N_i - \partial_i A), \\
\delta g_{ij} &=& a^2 [-h_{ij} - \partial_iQ_j - \partial_jQ_i + 2(\psi \delta_{ij} - \partial_i \partial_j E)], \nonumber
\end{eqnarray}
where  $\varphi$, $\psi$, A and E are scalar fields, $N_i$ and $Q_i$ are transverse vector fields and $h_{ij}$ is the transverse-traceless tensor perturbation.

 By expanding $\sqrt{-g + \delta g}$, $X(g + \delta g)$, $V^i(g + \delta g)$, $W^{ij}(g + \delta g)$ in  Eq.(\ref{ug}) and using Eq.(\ref{action}) we get the Lagrangian as
\begin{equation}
L_m = \frac{m_{pl}^2}{2}(m_0^2h_{00}h_{00} + 2m_1^2h_{0i}h_{0i} - m_2^2h_{ij}h_{ij} + m_3^2h_{ii}h_{jj} - 2m_4^2h_{00}h_{ii}),
\end{equation}
where the mass parameters  are given by \cite{mvp},
\begin{eqnarray}
m_0^2 &=& \frac{\Lambda^4}{m_{pl}^2}[XF_X + 2X^2 F_{XX}], \nonumber \\
m_1^2 &=& \frac{2 \Lambda^4}{m_{pl}^2}[-XF_X -WF_W + \frac{1}{2}XWF_{VV}], \nonumber \\
m_2^2 &=& \frac{2 \Lambda^4}{m_{pl}^2}[WF_W - 2W^2F_{WW2}], \\
m_3^2 &=& \frac{\Lambda^4}{m_{pl}^2}[WF_W + 2W^2F_{WW1}], \nonumber \\
m_4^2 &=& -\frac{\Lambda^4}{m_{pl}^2}[XF_X + 2XWF_{XW}], \nonumber
\end{eqnarray}
where
\begin{eqnarray}
W &=& -1/3 \delta_{ij} W^{ij}, \nonumber \\
\frac{\partial F}{\partial X} &=& F_X, \nonumber \\
\frac{\partial^2 F}{\partial X^2} &=& F_{XX}, \nonumber \\
\frac{\partial F}{\partial W^{ij}} &=& F_W\delta_{ij}, \\
\frac{\partial^2 F}{\partial V^i \partial V^j} &=& F_{VV}\delta_{ij}, \nonumber \\
\frac{\partial^2 F}{\partial W^{ij} \partial W^{kl}} &=& F_{WW1}\delta_{ij}\delta_{kl} + F_{WW2}(\delta_{ik}\delta_{jl} +\delta_{ij}\delta_{jk}), \nonumber \\
\frac{\partial^2 F}{\partial X \partial W^{ij}} &=& F_{XW}\delta_{ij}. \nonumber
\end{eqnarray}

There is a number of different regions in the mass parameter space where massive gravity is described by a consistent low-energy effective theory with strong coupling scale $\Lambda \sim (m m_{pl})^{1/2}$ which implies a ghost-free scenario. Each of these regions is characterized by certain fine-tuning relations between the mass parameters. In the vector sector, provided $m_2 \neq 0$, the vector field behaves in the same way as in the Einstein theory in the gauge $F_i = 0$; hence there are no propagating vector perturbations and gravity is not modified in this sector unless one takes into account the non-linear effects or higher derivative terms   \cite{Rubakov, Dubovsky}. In the scalar sector, the scalar field has massless limit which coincides with the GR expression; hence there is no vDVZ discontinuity. In the tensor sector, only the transverse-traceless perturbations $h_{ij}$ are present and their field equation is that of a massive field with the mass $m_2$ with helicity-2; hence there are two massive spin-2 propagating degrees of freedom.

The perturbed metric for  a flat FLRW universe can  be written as
\begin{equation}
ds^2 = a^2(\tau) [-d\tau^2 + (\delta_{ij} + h_{ij}) dx^i dx^j],
\end{equation}
here $\delta_{ij}$ is the flat space metric and $\tau$ is the conformal time defined by $d\tau = \frac{dt}{a}$.

The dynamical equation of motion for massive gravitational waves can be written as
\begin{equation}\label{emij}
h_{ij}^{(m) \prime \prime}(\tau) + 2\mathcal{H}h_{ij}^{(m)\prime}(\tau) + k^2h_{ij}^{(m)}(\tau) + a^2 m_{gw}^2 h_{ij}^{(m)}(\tau)  = 0,
\end{equation}
where $m_{gw} \equiv m_2$ is the mass of the graviton  and  $\mathcal{H} = \frac{a'}{a}$ is the Hubble parameter.

The massive tensor perturbation $h_{ij}^{(m)}$ can be expanded in the Fourier space as
\begin{eqnarray}\label{fmm}
\nonumber h_{ij}^{(m)} (\textbf{x}, \tau) &=& \frac{D}{(2\pi)^{\frac{3}{2}}} \int^{\infty}_{-\infty} \frac{d^3\textbf{k}}{\sqrt{2E_k}} [h_{ij}^{(m)(p)} (\tau) c_{ij}^{(m)(p)} \varepsilon_{ij}^{(m)(p)} (\textbf{k}) e^{i\textbf{k}.\textbf{x}}  \\
&&+ h_{ij}^{(m)(p) \ast} (\tau) c_{ij}^{(m)(p) \dagger} \varepsilon_{ij}^{(m)(p) \ast} (\textbf{k}) e^{-i\textbf{k}.\textbf{x}}],
\end{eqnarray}
where $D = \sqrt{16\pi}l_{pl}$ is the normalization constant, $l_{pl}$
 is  the Planck length, $E_k$ is the energy of the mode, $(p)$ is the polarization index and the superscript $(m)$ stands for  the massive tensor perturbation.

The two polarization states $\varepsilon_{ij}^{(p)}$, $p=1,2$ are symmetric and transverse-traceless and satisfy the conditions
\[\varepsilon_{ij}^{(p)}\delta^{ij}=0, ~~\varepsilon_{ij}^{(p)}k^i=0, ~~\varepsilon_{ij}^{(p)}\varepsilon^{(p') ij} = 2\delta_{pp'}, ~~\varepsilon_{ij}^{(p)}(\textbf{-k})=\varepsilon_{ij}^{(p)}(\textbf{k}).\]
These polarizations are linear and are called the plus $(+)$ polarization and cross $(\times)$ polarization. 

The creation and annihilation operators $c_k^{(p) \dagger}$ and $c_k^{(p)}$ satisfy the  following relations
\begin{eqnarray}
\left[c_k^{(p)},c_{k'}^{(p') \dagger}\right]&=&\delta_{pp'}\delta^3(k-k'),\\
\left[c_k^{(p)},c_{k'}^{(p')}\right]&=&\left[c_k^{(p) \dagger},c_{k'}^{(p') \dagger}\right]=0.
\end{eqnarray}

Using  Eq.(\ref{fmm})  in Eq.(\ref{emij}), we get
\begin{equation}\label{emk}
h_k^{(m)\prime \prime} (\tau) + 2\mathcal{H} h_k^{(m)\prime} (\tau) + (k^2 + a^2m_{gw}^2) h_k^{(m)} (\tau) = 0.
\end{equation}
Here after we drop  the polarization index $(p)$ and the index $(m)$ for notational convenience.

The mode function  can be taken in the following form
\begin{equation}\label{mf}
\mu_k (\tau) = a (\tau) h_k (\tau).
\end{equation}
Using  Eq.(\ref{mf})  in Eq.(\ref{emk}) we get 
\begin{equation}\label{mfm}
\mu_k^{\prime \prime} + \left(k^2 + a^2 m_{gw}^2 -\frac{a''}{a} \right) \mu_k = 0.
\end{equation}
The dispersion relation can be written as \cite{gum}
\begin{equation}\label{dr}
\frac{k^2}{a^2} + m_{gw}^2 = w^2,
\end{equation}
where $w$ is known as the effective frequency. 

For the adiabatic vacuum, Eq.(\ref{emk}) has the solution
\begin{equation}\label{av}
h_k (\tau) \propto e^{-iwa\tau}.
\end{equation}

For  the frequency  lower than the rate of cosmic expansion, $w^2 \ll \mathcal{H}^2$, the mode is  termed  super-horizon mode. The tensor amplitudes are frozen and the mode stays outside the horizon and is constant, and its absolute value is
\begin{equation}\label{sb}
|h_k| = \mathcal{A}_{ex}(k), ~~~~\tau < \tau_{k},
\end{equation}
where $\mathcal{A}_{ex} (k) = \frac{\mathcal{H}_{ex}}{m_{pl} k^{3/2}}$, is the amplitude of the mode at the time of its generation and  $\mathcal{H}_{ex}$ is the expansion rate at the time of horizon exit during inflation,  $\tau_{k}$ is the time of horizon re-entry and $m_{pl}$ is the reduced Planck mass.

When $w$ is comparable to the rate of cosmic expansion, $w^2 \simeq \mathcal{H}^2$, for a mode with comoving momentum $k$, the corresponding time is called horizon crossing time. Assuming that the horizon re-entry takes place sufficiently rapidly, i.e., $\tau \simeq \tau_{k}$, then Eq.(\ref{av}) can be rewritten as 
\begin{equation}\label{hre}
h_k (\tau) = \frac{\mathcal{C}(k)}{\sqrt{w_{k} a_{k}^3}} e^{-iwa\tau}, ~~~~\tau \simeq \tau_{k},
\end{equation}
where $w_{k} \equiv w(\tau_{k}) = \mathcal{H}_{k}$ indicates horizon re-entry and  $\mathcal{C}(k)$ is a constant of integration.

With the evolution of the universe, the modes re-enter the horizon and their amplitudes are no longer constant. The frequency becomes higher than the rate of cosmic expansion, $w^2 \gg \mathcal{H}^2$, called sub-horizon mode. Once the mode re-enters the horizon, it oscillates. Its solution is given by Eq.(\ref{av})
\begin{equation}\label{sp}
h_k (\tau) = \frac{\mathcal{C}(k)}{\sqrt{w(\tau) a^3(\tau)}} e^{-iwa\tau}, ~~~~\tau > \tau_{k}.
\end{equation}

Using Eq.(\ref{sb}), Eq.(\ref{hre}) and Eq.(\ref{sp}), we get
\begin{equation}\label{sm}
\frac{|h_k(\tau)|}{\mathcal{A}_{ex}(k)} = \sqrt{\frac{w_{k}}{w(\tau)}\frac{a_{k}^3}{a^3(\tau)}}, ~~~~\tau > \tau_{k}.
\end{equation}
Replacing $w$ by $k/a$ and $\tau_{k}$ by $\tau_{k}^{GR}$, $GR$ indicating the massless case, we get the corresponding solution in the massless case as
\begin{equation}\label{sml}
\frac{|h_k^{GR}(\tau)|}{\mathcal{A}_{ex}(k)} = \frac{a_{k}^{GR}}{a(\tau)}, ~~~~\tau > \tau_{k}^{GR}.
\end{equation}
The two-point correlation function  for the massive gravitational waves can be written as
\begin{equation}
P(w_0) \equiv \frac{d}{d \ln w_0} \langle 0| h_{ij} h^{ij} |0 \rangle,
\end{equation}
where 
\begin{equation}
\langle 0| h_{ij} (\textbf{x},\tau) h^{ij} (\textbf{x},\tau) |0 \rangle = \frac{D^2}{2\pi^2} \int^{\infty}_0 k^2 |h_k (\tau)|^2 \frac{dk}{k}.
\end{equation}

Therefore one gets
\begin{equation}
P(w_0) = \frac{w_0^2}{w_0^2 - m_{gw,0}^2} \frac{2k^3}{\pi^2}|h_k(\tau_0)|^2,
\end{equation}
where 
\begin{eqnarray}
k = a_0\sqrt{w_0^2 - m_{gw,0}^2}, \nonumber \\
\frac{d}{d \ln w_0}\left(\frac{dk}{k}\right) = \frac{w_0^2}{w_0^2 - m_{gw,0}^2}.\nonumber
\end{eqnarray}

Using Eq.(\ref{sm}), the  power spectrum for the massive gravitational waves  is obtained as
\begin{eqnarray}\label{pm}
P(w_0) &=& \frac{2k^3}{\pi^2} \mathcal{A}^2(k) \left(\frac{k'a_{k}}{ka_0}\right)^2 \frac{w_{k}a_{k}}{w_0a_0} \nonumber \\
	&=& \left(\frac{k'a_{k}}{ka_0}\right)^2 \frac{w_{k}a_{k}}{w_0a_0} P(k),
\end{eqnarray}
where $k' = a_0w_0$ and $P(k) = \frac{2k^3}{\pi^2} \mathcal{A}^2(k)$ is known as the primordial power spectrum.

Using Eq.(\ref{sml}), the power spectrum for the massless case can be written as
\begin{equation}\label{pml}
P_{GR} (w_0) = \left(\frac{a_{k'}^{GR}}{a_0} \right)^2 P(k').
\end{equation}
By taking the ratio of Eq.(\ref{pm}) to Eq.(\ref{pml}), we obtain
\begin{eqnarray}
\frac{P(w_0)}{P_{GR}(w_0)} &=& \frac{P(k)}{P(k')} \left(\frac{k'a_k}{ka_{k'}^{GR}}\right)^2 \frac{w_ka_k}{w_0a_0} \nonumber \\
	&=& \frac{P(k)}{P(k')} S^2(w_0),
\end{eqnarray}
where the enhancement factor  $ S(w_0)$ can be written as
\begin{equation}\label{ef}
S(w_0) = \frac{k'a_k}{ka_{k'}^{GR}} \sqrt{\frac{w_ka_k}{w_0a_0}}.
\end{equation}

 The dispersion relation  at the time of horizon re-entry is
\begin{equation}
w_k \simeq m_{gw}(\tau_{k}).
\end{equation}

The cosmic expansion rate is comparable to the effective mass of the gravitational waves when all modes re-enter the horizon simultaneously, then
\[\mathcal{H}(\tau_{k}) \simeq m_{gw}(\tau_{k}).\]

Therefore, we have   $\eta_{k} \simeq \eta_{hc}$, $a_{k} \simeq a_{hc}$, $\mathcal{H}_{k} \simeq \mathcal{H}$ and $w_{hc} \simeq m_{gw}(\tau_{hc}) = \frac{k_{hc}}{a_{hc}}$.

By considering  the mass term which dominates the frequency modes till present time, we get
\begin{eqnarray}
w_0 \simeq m_{gw,0} = \frac{k_0}{a_0}, \nonumber \\
k' \simeq k_0. \nonumber
\end{eqnarray}
  For long wavelength modes,  the enhancement factor becomes \cite{gum}
\begin{equation}\label{eh}
S(w_0) \simeq \frac{a_{hc}}{a_{k_0}^{GR}}\sqrt{\frac{k_{hc}}{k_0}}\left(\frac{w_0^2}{m_{gw,0}^2}-1\right)^{-\frac{1}{2}}.
\end{equation}

The massive short wavelength modes behave almost similar to their massless counterparts and hence, are not considered here.

\section{Inflation}
In the simplest inflationary scenario, the sudden  expansion of  early universe is driven by a canonical single  scalar field called the inflaton. In the slow roll inflationary  scenario, the inflaton slowly rolls down its potential which is almost flat. 

The equation of motion for the inflaton with effective potential $V$ can be written as
\begin{equation}
\ddot{\phi} + 3H \dot{\phi} + V'(\phi) =0,
\end{equation}
where  the Hubble parameter $H$    
 is determined by the energy density of the inflaton  field,
 \[ \rho_{\phi} = \frac{\dot{\phi}^2}{2}+V,\] so that the Friedmann equation can be written as
\begin{equation}\label{fdmeq}
H^2 = \frac{1}{3 m^2_{pl}}\left( \frac{1}{2} \dot{\phi}^2 + V(\phi)\right).
\end{equation}
In the slow-roll limit, 
the Hubble parameter takes the following form 
\begin{equation}
H^2 \simeq \frac{V}{3m^2_{pl}}.
\end{equation}
 The slow-roll condition is characterized in terms of  the slow-roll parameters  defined in terms of the inflaton potential  and its derivatives as  follows
\begin{eqnarray}\label{sl}
\epsilon &\equiv & \frac{m_{pl}^2}{2}\left(\frac{V'}{V}\right)^2 \ll 1, \nonumber \\ 
\eta &\equiv &  m^2_{pl}\left(\frac{V''}{V}\right) \ll 1,
\end{eqnarray}
these are  the sufficient but not necessary conditions. As long as the slow-roll conditions are satisfied the  inflation continues. The slow-roll approximation can be used to study the fluctuations generated during inflation.

The strength of the tensor fluctuations can be measured with respect to that of the scalar fluctuations  in terms of the tensor-to-scalar ratio $r$ given by
\begin{equation}\label{r}
r \equiv \frac{P_T (k)}{P_S (k)} \simeq 16\epsilon.
\end{equation}

The tensor spectral index can be given by the parameter $\epsilon$ as
\begin{equation}\label{nT}
n_T = -2\epsilon.
\end{equation}

It can be seen  that both $r$ and $n_T$ are determined by the equation of state during inflation. They are  useful in understanding the dynamics of the early universe as well as  in distinguishing various  inflationary  models. 

\subsection{Inflation models}
In this work, we consider  the single field slow-roll inflation models for which  
the corresponding tensor-to scalar ratio  lies within 
 $\mathcal{O}(10^{-3})$  and $r < 0.07$ \cite{gb}. The scalar power spectrum for each model is taken to be $P_S = 2.43 \times 10^{-9}$.

\subsubsection*{R2 Inflation model (Starobinsky model)}
This model is based on the higher order gravitational terms with the action \cite{st}
\begin{equation}
S = \int d^4x \sqrt{-g}\frac{m_{pl}^2}{2}\left(R+\frac{R^2}{6m^2}\right),
\end{equation}
where $R$ is the Ricci scalar and $m$ is the inflaton mass.

The model can be represented in the form of Einstein gravity with a normalized inflaton field with effective potential,
\begin{equation}
V(\phi) = M^4 (1-e^{-\sqrt{2/3}\phi/m_{pl}})^2.
\end{equation}
The tensor-to-scalar ratio for this model is obtained as $r = 3.25 \times 10^{-3}$.
The slow-roll parameters  obtained for the model  are 
\begin{eqnarray}
\epsilon &=& 2.03 \times 10^{-4}, \nonumber \\
\eta &=& -1.63 \times 10^{-2}. 
\end{eqnarray}

The calculated  tensor power spectrum with the tensor spectral index $n_T = -4.06 \times 10^{-4}$  is  $P_T = 7.9 \times 10^{-12}.$
 
\subsubsection*{Arctan Inflation model}
This model is considered as a large field inflation where the inflaton field starts at a large value and then evolves to the minimum potential \cite{ai1, ai2}. The effective potential for this model is given by
\begin{equation}
V(\phi) = M^4 \left[1 - \arctan\left(\frac{\phi}{\mu}\right)\right],
\end{equation}
where $\mu/m_{pl} = 10^{-2}$ is a free parameter which characterizes the typical vacuum expectation value at which inflation takes place, $M/m_{pl} = 10^{-3}$.

The tensor-to-scalar ratio for this model is found  as $r = 1.38 \times 10^{-2}$.
The calculated slow-roll parameters are,
\begin{eqnarray}
\epsilon &=& 8.62 \times 10^{-4}, \nonumber \\
\eta &=& 3.0 \times 10^{-2}. 
\end{eqnarray}

The obtained tensor power spectrum is  $P_T = 3.35 \times 10^{-11}$ for which  the tensor spectral index has the value  $n_T = -1.72 \times 10^{-3}$.

\subsubsection*{Higgs Inflation model}
In this model, the Higgs field is considered to play the role of the inflaton. The field is considered to be non-minimally coupled to gravity \cite{higgs}. The effective potential for this model is
\begin{equation}
V(\phi) = M^4 (1 + e^{-\sqrt{2/3}\phi/m_{pl}})^{-2}.
\end{equation}
The tensor-to-scalar ratio for this model is  $r = 2.83 \times 10^{-3}$.
The corresponding slow-roll parameters are,
\begin{eqnarray}
\epsilon &=& 1.77 \times 10^{-4}, \nonumber \\
\eta &=& -1.48 \times 10^{-2}. 
\end{eqnarray}

The tensor power spectrum is  obtained as $P_T = 6.87 \times 10^{-12}$ with  $n_T = -3.53 \times 10^{-4}$.

\subsubsection*{Inverse Monomial Inflation model}
This model is considered in the context of quintessential inflation where the inflaton need not necessarily decay and hence, may survive through the present epoch. Since the inflaton does not decay, radiation is created via gravitational particle production \cite{im1, im2, im3}. The effective potential for this model is
\begin{equation}
V(\phi) = M^4 \left(\frac{\phi}{m_{pl}}\right)^{-p},
\end{equation}
where $p$ is a positive parameter, $M/m_{pl} = 10^{-1}$.

The calculated  tensor-to-scalar ratio for this model  is   $r = 2.0 \times 10^{-3}$ and 
the slow-roll parameters are,
\begin{eqnarray}
\epsilon &=& 1.25 \times 10^{-4}, \nonumber \\
\eta &=& 3.33 \times 10^{-4}. 
\end{eqnarray}

The tensor power spectrum is found as  $P_T = 4.86 \times 10^{-12}$ with  $n_T = -2.50 \times 10^{-4}$.

\subsubsection*{Loop Inflation model}
This model is studied in the context of spontaneous symmetry breaking which alters the flatness of the potential and takes the form of logarithmic function for one loop order correction \cite{li1, li2, li3}. The effective potential for this model is
\begin{equation}
V(\phi) = M^4 \left[1 + \alpha \ln \left(\frac{\phi}{m_{pl}}\right)\right],
\end{equation}
where $\alpha = g^2/16\pi^2$ tunes the strength of radiative effects, $M = 10^{16}$ GeV.

The tensor-to-scalar ratio for this model is obtained as  $r = 4.34 \times 10^{-2}$.
The calculated slow-roll parameters are,
\begin{eqnarray}
\epsilon &=& 3.09 \times 10^{-3}, \nonumber \\
\eta &=& -2.06 \times 10^{-2}. 
\end{eqnarray}

The  calculated tensor power spectrum with the tensor spectral index $n_T = -6.18 \times 10^{-3}$ is  $P_T = 1.2 \times 10^{-10}.$

\section{The BB-mode angular power spectrum  of CMB}
The expression for computing the  $BB$-mode correlation  angular power spectrum of  CMB is given by \cite{cmb4, cmb5} 
\begin{eqnarray}
C_l^{BB} &=& (4\pi)^2 \int dk k^2 P_T(k) \nonumber \\ && \times \left| \int_0^{\tau_H} d\eta g(\tau) h_k(\tau) \Big\{(8x + 2x^2 \partial_x)\frac{j_l(x)}{x^2}\Big\}_{x=k(\tau_0-\tau)}\right|^2,
\end{eqnarray}
where $x= k(\tau_0 -\tau)$, $g(\tau) = \kappa e^{-\kappa}$  is  the probability distribution of the last scattering with  $\kappa$ as  the differential optical depth  and $j_l(x)$ is the spherical Bessel function.

\begin{figure}
\includegraphics[scale=0.35]{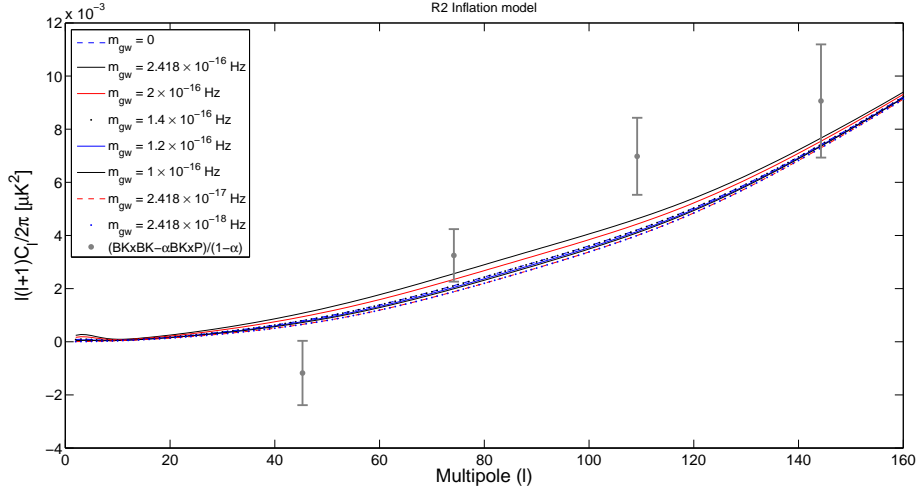}
\caption{Lensed BB-mode correlation  angular spectrum of CMB for the Starobinsky (R2) inflation model  for various values of graviton mass  with  the   BICEP2/Keck Array  and Planck  joint data.}\label{f1}
\end{figure}

\begin{figure}
\includegraphics[scale=0.35]{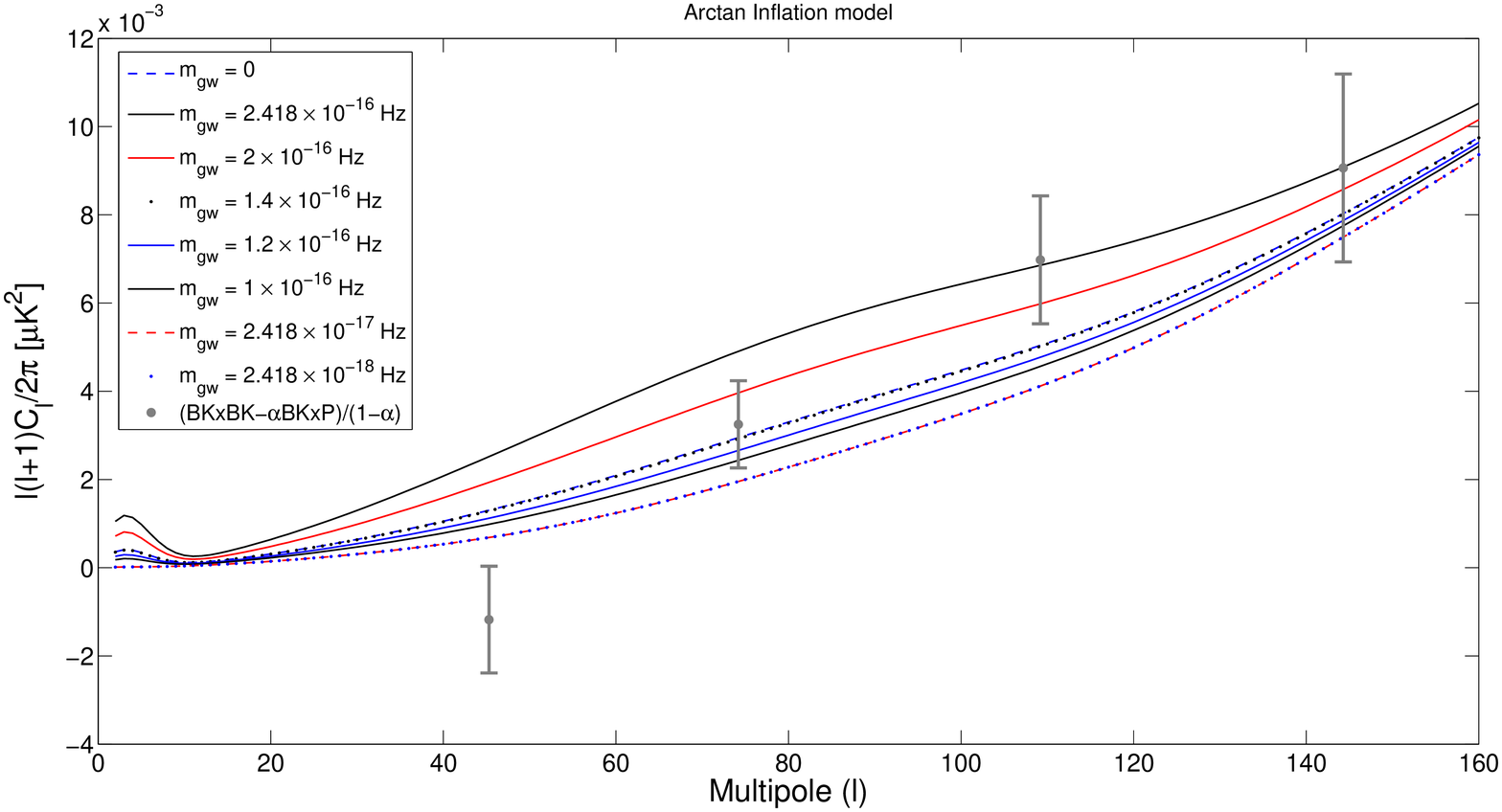}
\caption{Lensed BB-mode correlation  angular spectrum of CMB for the Arctan inflation model for various values of graviton mass  with  the   BICEP2/Keck Array  and Planck  joint  data.}\label{f2}
\end{figure}

\begin{figure}
\includegraphics[scale=0.35]{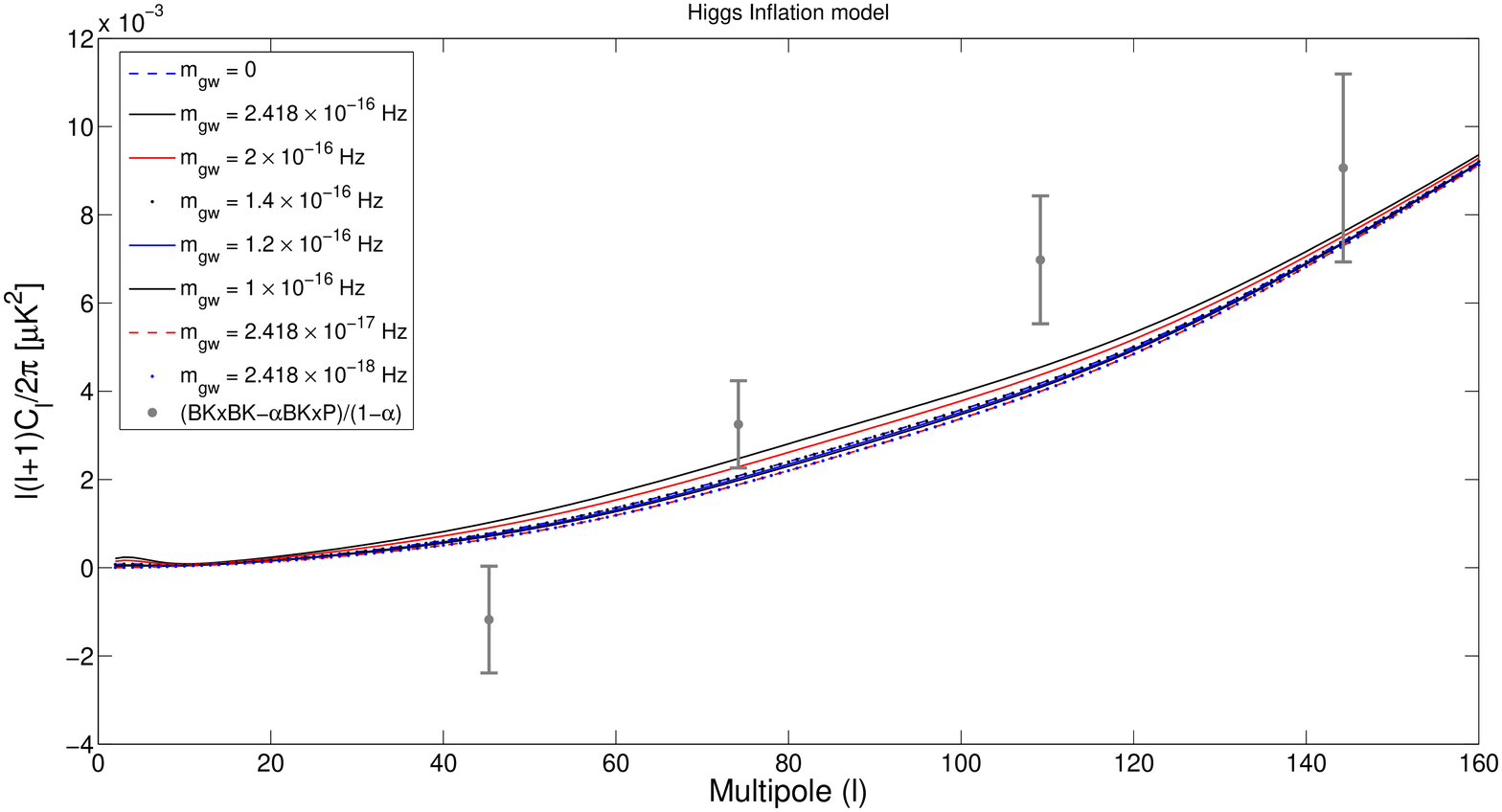}
\caption{Lensed BB-mode correlation  angular spectrum of CMB for the  Higgs inflation model  for various values of graviton mass with  the   BICEP2/Keck Array  and Planck  joint data.}\label{f3}
\end{figure}

\begin{figure}
\includegraphics[scale=0.35]{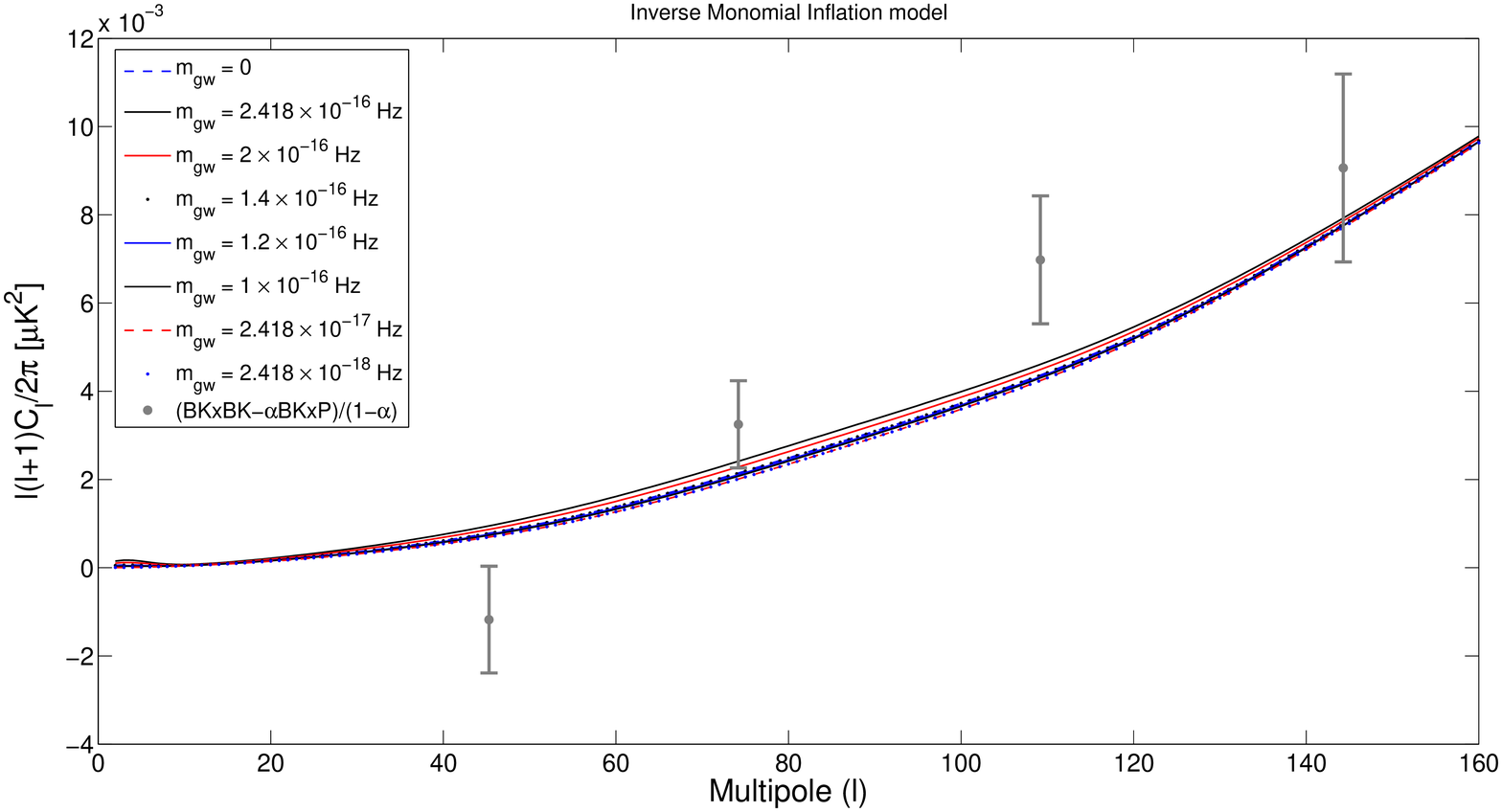}
\caption{Lensed BB-mode correlation  angular spectrum of CMB for  the Inverse monomial inflation model  for various values of graviton mass with  the   BICEP2/Keck Array  and Planck  joint  data.}\label{f4}
\end{figure}

\begin{figure}
\includegraphics[scale=0.35]{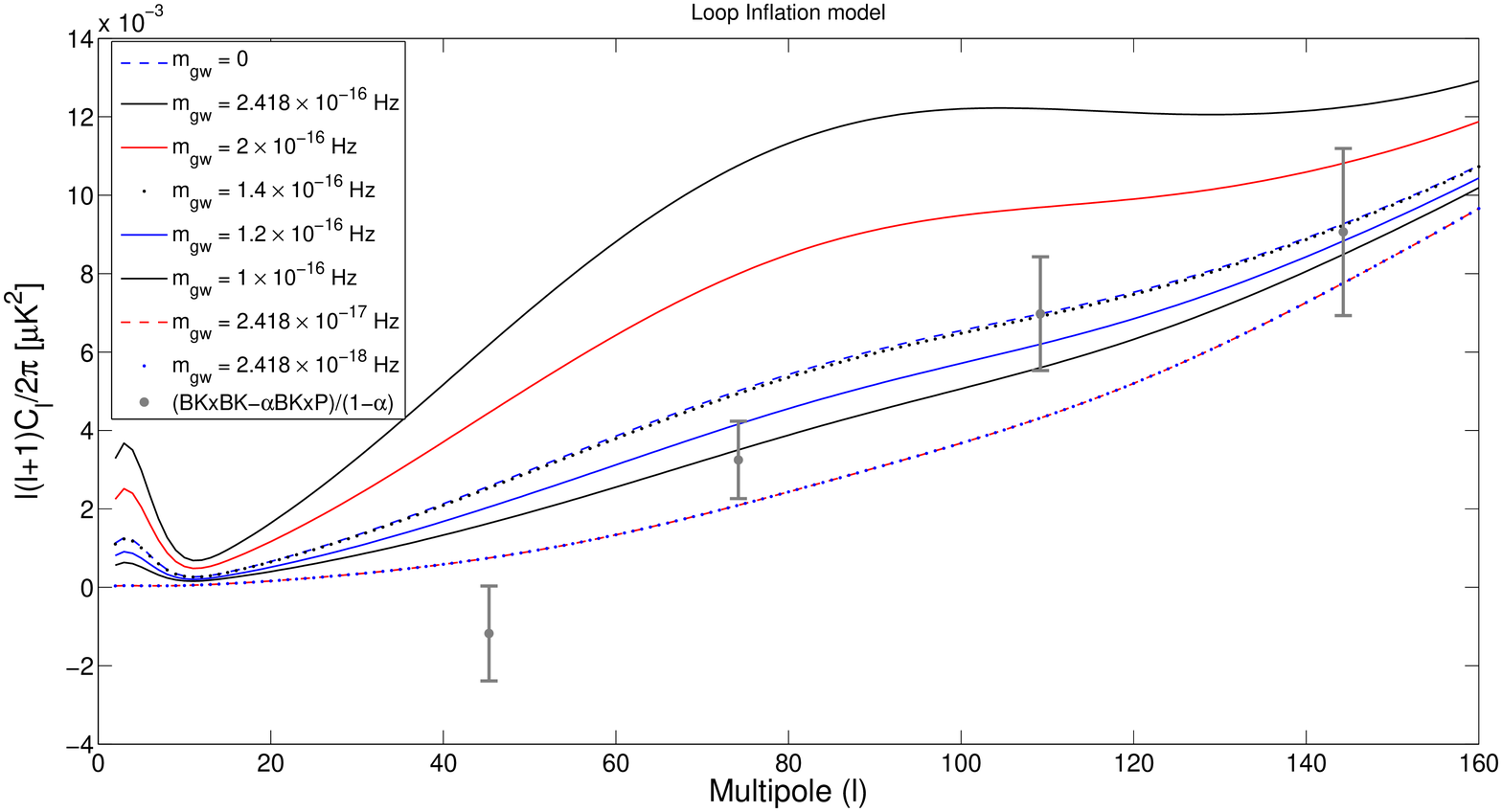}
\caption{Lensed BB-mode correlation  angular spectrum of CMB  for the Loop inflation model  for various values of graviton mass with  the   BICEP2/Keck Array  and Planck  joint data.}\label{f5}
\end{figure}

The CMB angular spectrum for the  BB mode correlation with the slow-roll inflation models are obtained  by using the CAMB code with  $\kappa = 0.08$  and $k_0 = 0.002$  Mpc$^{-1}$ as the tensor pivot scale.  The obtained results are presented in 
Figs.\ref{f1}, \ref{f2}, \ref{f3}, \ref{f4} and \ref{f5}.   The limit $(BK \times BK - \alpha BK \times P)/(1 - \alpha)$ is taken  from the BKP joint data after subtraction of dust contribution of the BICEP2/Keck Array band which gives the fiducial value $\alpha = 0.04$ \cite{bkp}.  

\section{Discussion and conclusion }
The BB mode correlation angular power spectrum of CMB for  the primordial massive gravitational waves   for  the Starobinsky (R2), Arctan, Higgs, Inverse monomial and Loop  inflation models is studied in the context of Lorentz violating massive gravity model. It is observed for   each inflation  model that,  for gravitational waves with mass $m_{gw} \gtrsim 1.4 \times 10^{-16}$ Hz, there is enhancement in the power level compared to the massless gravitational waves case while  there is  decrease in the   power level  in the case of  $m_{gw} < 1.4 \times 10^{-16}$ Hz. The BB mode angular power spectrum of CMB for gravitational waves with mass $m_{gw} \simeq 1.4 \times 10^{-16}$ Hz ($\equiv 5.79 \times 10^{-31}$ eV) is found almost comparable  to  its  massless counterpart.  The increase/decrease in the power level of  BB mode angular power spectrum of CMB  for the massive gravitational waves   is greater for inflation models with larger deviation ($n_T$) from scale invariance. For each slow-roll inflation model, the  angular power spectrum  for  the  gravitational waves  with masses $m_{gw} = 2.418 \times 10^{-17}$ Hz ($\equiv 10^{-31}$ eV) and $m_{gw} = 2.418 \times 10^{-18}$ Hz ($\equiv 10^{-32}$ eV) are found marginally within the limit  of  BICEP2 and Planck joint data at higher multipoles, which indicates  that the lower limit for the graviton mass may be higher than these masses. At the same time, the upper limit for the primordial graviton mass may also be higher than $m_{gw} = 10^{-30}$ eV. Hence, the results and analysis of the present study  on the  BB mode angular power spectrum of CMB with the  BICEP2/Keck Array and Planck joint data  for various inflationary models show that the mass limit for primordial graviton may be higher than  the earlier proposals \cite{Dubovsky,cmbmg2,mgtensor}. The present study may be repeated with  other inflation models  which does not seem to  alter the conclusions of the present study and  is currently beyond the scope of the present work.

\section*{Acknowledgement}
N M would like to thank Council for Scientific and Industrial Research (CSIR), New Delhi for financial support. PKS  thank SERB, New Delhi for the  financial support.

%\section*{Reference}

\end{document}